\documentclass[twocolumn]{aastex63}
\received{XXXX}
\revised{YYYY}
\accepted{ZZZZ}
\submitjournal{ApJ}

\shorttitle{Dust growth in molecular cloud envelopes}
\shortauthors{Beitia-Antero and G\'omez de Castro}

\begin{document}

\title{Dust growth in molecular cloud envelopes: a numerical
approach}

\correspondingauthor{Leire Beitia-Antero}
\email{lbeitia@ucm.es}

\author[0000-0003-0833-4075]{Leire Beitia-Antero}
\affiliation{Joint Center for Ultraviolet Astronomy \\
Universidad Complutense de Madrid \\
Edificio Fisac (Fac. Estudios Estad\'isticos), Av. Puerta de Hierro s/n, 28040, Madrid, Spain}
\affiliation{Departamento de F\'isica de la Tierra y Astrof\'isica (U.D. Astronom\'ia y Geodesia)\\
Facultad de CC. Matem\'aticas, Universidad Complutense de Madrid \\
Plaza de las Ciencias 3, 28040, Madrid, Spain}

\author[0000-0002-3598-9643]{Ana I. G\'omez de Castro}
\affiliation{Joint Center for Ultraviolet Astronomy \\
Universidad Complutense de Madrid \\
Edificio Fisac (Fac. Estudios Estad\'isticos), Av. Puerta de Hierro s/n, 28040, Madrid, Spain}
\affiliation{Departamento de F\'isica de la Tierra y Astrof\'isica (U.D. Astronom\'ia y Geodesia)\\
Facultad de CC. Matem\'aticas, Universidad Complutense de Madrid \\
Plaza de las Ciencias 3, 28040, Madrid, Spain}



\begin{abstract}
  Variations in the grain size distribution are to be expected
  in the interstellar medium (ISM) due to grain growth and destruction.
  In this work, we present a dust collision model to be implemented
  inside a magnetohydrodynamical (MHD) code that takes into account
  grain growth and shattering of charged dust grains of a given
  composition (silicate or graphite). We integrate this model
  in the MHD code Athena, and builds on a previous implementation
  of the dynamics of charged dust grains in the same code. To
  demonstrate the performance of this coagulation model, we
  study the variations in the grain size distribution of a single-sized
  population of dust with radius 0.05 $\mu$m inside several dust
  filaments formed during a 2D MHD simulation. We also
  consider a realistic
   dust distribution with sizes ranging from 50 \AA~to 0.25 $\mu$m
  and analyze both the variations in the size distribution for
  graphite and silicates, as well as of the far ultraviolet
  extinction curve. From the obtained results, we conclude
    that the methodology here presented, based on the MHD evolution
    of the equation of motion for a charged  particle, is optimal for studying
    the coagulation of charged dust grains in a diffuse regime
    such as a molecular cloud envelope. Observationally, these
    variations in the dust size distribution are translated into
    variations in the far ultraviolet extinction curve, and they
  are mainly caused by small graphite dust grains.

\end{abstract}

\keywords{Interstellar medium -- Interstellar dust -- Ultraviolet extinction}


% ------------------------------------- %
% ------------------------------------- %
\section{Introduction} \label{sec:intro}
% ------------------------------------- %
% ------------------------------------- %
Dust grains are ubiquitous in space. They are formed
in the atmospheres of evolved stars \citep{1989A&A...223..227D}
or even in supernova (SN) shocks \citep{2003ApJ...598..785N}, and play
a fundamental role in the evolution of the interstellar
medium (ISM). They are one of the main
sites of molecule formation \citep{1971ApJ...163..155H}
and constitute one of the principal heating mechanisms of the
ISM \citep{1994ApJ...427..822B,2001ApJS..134..263W}. Besides,
dust grains may also affect the fragmentation and evolution
of molecular clouds, since in the diffuse phases of the ISM they
acquire a net charge that favors the coupling with
the magnetic field and
interfere with the propagation of magnetohydrodynamic (MHD) waves
\citep{1987ApJ...314..341P,2006MNRAS.371..513C,2019AJ....157...83P}.\par

Over their lifetime, interstellar dust grains go through several
growing and shattering phases that depend on the ambient conditions.
In a dense enough medium, two grains may collide and form
a larger particle if their relative velocity is lower
than a given threshold (\citealt{2009MNRAS.394.1061H},
hereafter HY09); if the velocity is larger,
then shattering takes place and a full distribution of dust
grains is generated \citep{2010A&A...513A..56G}. At very high gas temperatures
characteristic of the hot intracluster medium or of
SN shocks ($T\geq 10^{6}$ K) dust grains may be eroded
through collisions with thermally excited gas
\citep{1979ApJ...231...77D}. Finally,
dust grains may also grow by accretion of gas-phase metals
\citep{1998ApJ...501..643D,2016ApJ...831..147Z}.\par

Variations in the small dust population, especially in those grains
with a carbonaceous nature, produce noticeable variations in the
ultraviolet (UV) extinction curve. The main feature
is the so-called UV bump at 2175 \AA~that is caused by
very small carbonaceous particles; polycyclic aromatic
hydrocarbon (PAH) molecules have been put forward
as the main carriers of the bump \citep{2001ApJ...548..296W}
although there may be a substantial contribution of
small graphite grains \citep{1965ApJ...142.1681S}
or even multi-shell fullerenes, usually called buckyonions
\citep{2004ApJ...608L..37I}. Variations in the strength of the UV bump
have been observed in our Galaxy
\citep{1984ApJ...279..698W,2015MNRAS.449.3867G,2017MNRAS.469.2531B}
but also towards
other nearby galaxies \citep{1984A&A...132..389P,2019MNRAS.486..743D}. 
The other characteristic feature of the UV extinction curve is the
steep slope at far ultraviolet (FUV) wavelengths. This feature
is mainly due to the effective absorption and scattering of UV photons by
very small dust grains \citep{2012MNRAS.423.2941R}, and its
parametrization is simpler than that of the
bump \citep{2007ApJ...663..320F}.\par

In a previous work (\citealt{2020arXiv200813135B},
hereafter Paper I) we studied the formation
of dust filaments under conditions typical of a molecular cloud envelope
in 2D.
We considered passive dust grains of size
$a_{d} = 0.05~\mu$m and  negatively charged ($Z_{d} = -17$) that
evolved under the sole influence of gas and magnetic fields.
In this article, we explore the growth of charged dust inside
those filaments using a basic collision model
that we have implemented
in the MHD code Athena \citep{2008ApJS..178..137S}.
In Sec. \ref{sec:algorithm}, we describe
the  model and its integration in Athena.
Then, in Sec. \ref{sec:simu_turb}, we apply this module to
study dust growth in the filaments from Paper I.
In Sec. \ref{sec:simu_real} we perform a similar study
but considering a realistic dust size distribution, and
derive the predicted variations in the FUV slope of the
UV extinction curve. Finally, in Sec. \ref{sec:discussion} we
discuss the results and present the main conclusions.

% ------------------------------------- %
% ------------------------------------- %
\section{Dust collision model in Athena}\label{sec:algorithm}
% ------------------------------------- %
% ------------------------------------- %
Dust collision models consider different outcomes
based on the medium of interest. For instance, when
dealing with large ($a_{d} > 100~\mu$m) grains in
protoplanetary disks, the fragmentation of dust grains may
be very complex \citep{2010A&A...513A..57Z}, while in the ISM
shattering by
SN induced turbulence might be considered
\citep{2013EP&S...65..183H}. \par

We have developed a collision model to be
integrated inside a MHD code that follows
the dynamics of dust particles. This is fundamentally
different from the usual approach that evolves either
a dust size distribution \citep{2009MNRAS.394.1061H} or a sample of
test particles via Monte Carlo methods \citep{2008A&A...489..931Z} where
the dust particles have a prescribed velocity.
We allow the computational particles to evolve in 2D according
to their equation of motion as explained in Paper I, and
only evaluate their interactions at the end of every
time step. Therefore, we have taken into account
several constraints:

\begin{itemize}
\item Computationally, only a limited number of particles
  may be followed. This implies that all the dust mass will be
  distributed among a finite number of test particles.

\item After an erosion event, part of the mass
  of one of the test particles will
  be lost. We hereby suppose that this dust mass corresponds to
very small grains (sizes of a few \AA) that return to the gas phase.

\item A given particle may only interact once per time step, so
  even if more than two particles are near enough to interact, only
  two of them will be selected. For the simulations presented
    in Sec. \ref{sec:simu_turb}, typically 35\% of the interactions
    are subject to this multiplicity at the initial stage.
    However, this ambiguity is
    rapidly resolved by the system because nearby
    particles will interact with each other after a few computational
  steps.
  
\item Dust particles are either silicates or graphites, and only
  one particle species is allowed at the same time.
  
\end{itemize}

Our model
is built specifically for the study of dust
coagulation in the diffuse phases of the ISM, and
we have only contemplated two possible outcomes
for a collision between particles: they may grow
if their relative velocity is lower than a threshold, or
one of them (the smallest one) loses mass if their
  relative velocity is larger than the critical value,
  a phenomenon
that we refer to as erosion. In the following
section we provide a detailed explanation of the algorithm.
This algorithm has been developed within the 2D framework imposed
  by the simulations from Paper I. However, the philosophy behind the method is also applicable in 3D.

% % % % % % % % % % % % % % % % % % % % % % % % % % 
\subsection{Collision algorithm}
\begin{figure}[h!]
  \plotone{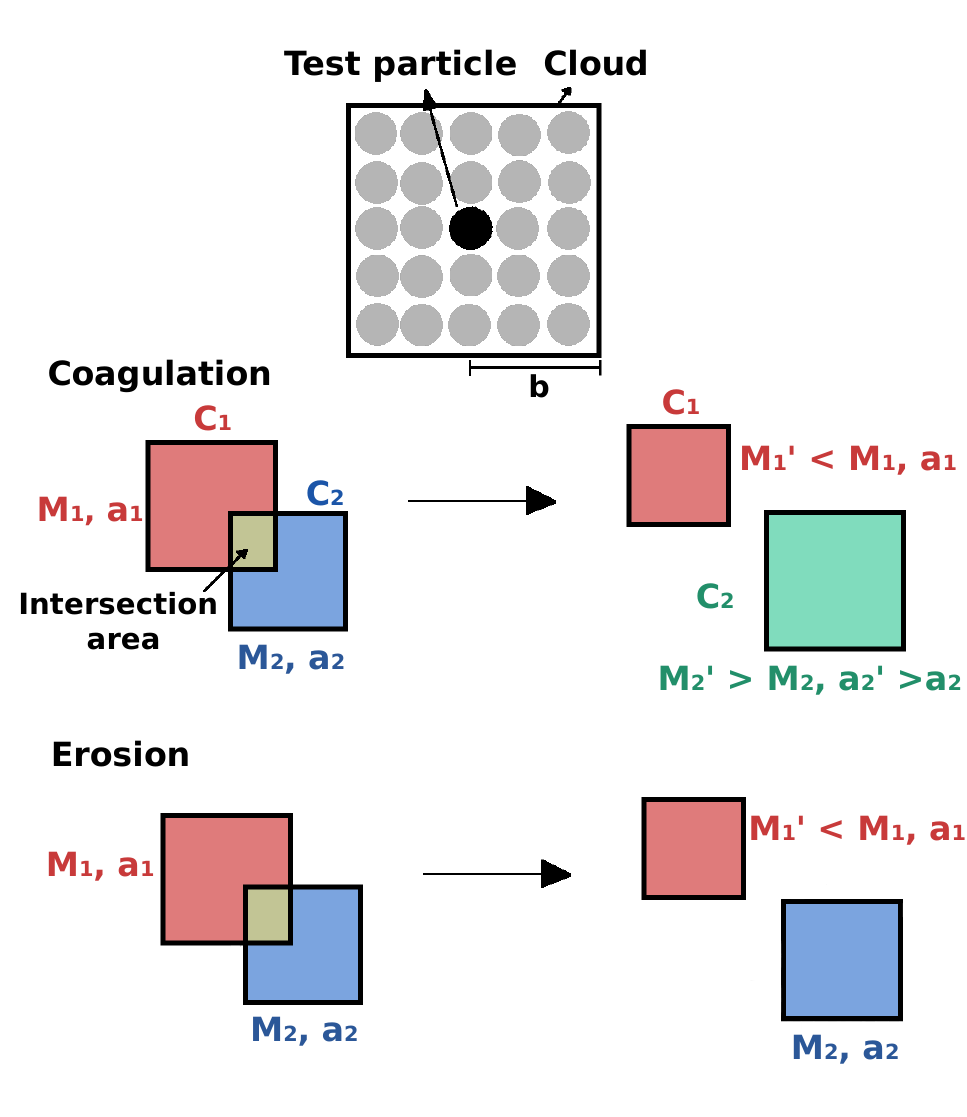}
  \caption{Illustration of the collision model implemented in Athena.
    At the top, we represent the shape of a test particle (black circle),
    referred to
    as `cloud'. Grey circles represent the real particles contained
    inside the cloud that are indirectly followed.
    In the middle we show the basic modifications introduced
    after a coagulation event, while the bottom of the image
    show the treatment of an erosion event.
    In the middle and bottom panels, the cloud shape is
      proportional to the represented dust mass and the color indicates
      the radius of the represented particles. Hence, two clouds of particles
      with different radii ($a_{1}$ in red, $a_{2}$ in blue)
      are considered initially,
      with $a_{2} > a_{1}$. After
      a coagulation event, the cloud of particles with larger radii (blue)
      increases
      its represented mass (larger cloud size) and its radius ($a_{2}' > a_{2}$,
      green color), while the smaller cloud (red) loses represented mass
      but maintains the particle radius. After an erosion
      event, the cloud of biggest particles (blue) keeps its properties
      while the smallest one (red) loses represented mass, but
    also maintain the particle radius.
  }
  \label{fig:geometry}
\end{figure}

% % % % % % % % % % % % % % % % % % % % % % % % % % 
In a given region, we have a total amount of dust
$M_{d}$ distributed over a given number of
particles $N_{par}$. In practice, $N_{par}$ is so large
that it is unfeasible to follow the
evolution of every single grain. Therefore, the
common approach is to consider a finite number
of test particles that represent a swarm of real
particles. We shall suppose that the behavior of
a test particle is representative of that of
$k$ real particles, so they will have the same
charge, velocity, and approximately the same
position in space. For the implementation of
the collision algorithm, we consider the computational
test particles as the centroids of clouds of real
particles uniformly distributed in a 2D domain
(see Fig. \ref{fig:geometry} for an illustration). In
order to avoid confusion, in the rest of this section we will
use the term \textit{clouds} for the test particles, and will
reserve the word \textit{particle} for referring
to each of the $N_{par}$ real particles.\par

The size of the clouds is imposed by the properties
of the particles they represent. If a cloud contains
$k$ particles of radius $a_{d}$, then its linear size is $2b$,
where $b = a_{d}\sqrt{k}$. This prescription is based on
the assumption that the cloud has a squared shape, which
corresponds to a uniform cartesian 2D distribution of the particles.
Then, two clouds $C_{1}$ and $C_{2}$ will interact if the
distance between them is $d < b_{1} + b_{2}$, and
the outcome of the interaction will be determined
by their relative velocity $v = |{\bf v_{1}} - {\bf v_{2}}|$:

\begin{itemize}
\item If $v > v_{crit}$, then the cloud of smaller particles
  loses mass (erosion).
\item If $v < v_{crit}$, grain growth takes place (coagulation). In that
  case, we suppose that smaller particles get stuck into the
  larger ones, and the properties of both clouds will be modified.
\end{itemize}

The critical velocity $v_{crit}$ depends on the grain material,
and must be set by the user. In our simulations, we have
adopted $v_{crit} = 2.7$ km s$^{-1}$ for silicates and
$v_{crit} = 1.2$ km s$^{-1}$ for graphites
as in HY09.\par

The properties of the clouds $C_{1}$ and $C_{2}$ will be modified
as a function of the intersection area between them, $A_{\rm inter}$.
After an erosion event, only the cloud of smaller particles, let us
say $C_{1}$, loses mass: $M_{1}' = M_{1} - A_{\rm inter}M_{1}/A_{1}$, where
$M_{1}$ is the mass of the cloud and $A_{1} = 4b^{2}$ its area.
For coagulation however, both clouds suffer complex modifications
that have to be taken into account:

\begin{itemize}
\item The cloud of smaller particles ($C_{1}$) loses mass
  as when an erosion event takes place.
  
\item Particles inside the $C_{2}$ cloud will experience growth.
  We suppose that there are $k_{1}^{\rm inter}$ and
  $k_{2}^{\rm inter}$ particles in the intersection area $A_{\rm inter}$,
  and the smaller particles get stuck into the bigger ones. Hence,
  in the intersection area, the radius of the particles will be
  $a_{2}^{\rm inter} > a_{2}$. The final radius of particles inside
  the $C_{2}$ cloud is computed through a weighted mean between
  the value inside the intersection area, $a_{2}^{\rm inter}$, and
  outside of it, $a_{2}$:

  \begin{equation}
    a_{2}' = a_{2}^{\rm inter}\frac{A_{\rm inter}}{A_{2}} +
    \bigg(1 - \frac{A_{\rm inter}}{A_{2}}\bigg)a_{2}
  \end{equation}

\item Since $C_{1}$ transfers mass to $C_{2}$, it is to expect
  that momentum is also transferred during the coagulation
  process. If we denote by $m_{1}$ and $m_{2}$ the masses of
  particles inside the clouds $C_{1}$ and $C_{2}$, then
  the  velocity of the latter species is computed as:

  \begin{equation}
    m_{2}'{\bf v}_{2}' = m_{2}{\bf v}_{2} + \varepsilon m_{1}{\bf v}_{1}
  \end{equation}

  where $\varepsilon = -v/v_{crit} + 1$ is an efficiency factor
  that accounts for the fraction of
  energy transferred, and
  $(1 - \varepsilon)$ is the energy lost
  in creating the chemical bonds of the new particles. For
  $\varepsilon  = 1$, perfect coagulation takes place,
  while for $\varepsilon = 0$ all the energy is lost and
  erosion becomes the dominant outcome.

\item Finally, the other particle properties such as charge ($Z_{d}$),
  Coulomb parameter $\nu_{C}$, and cloud linear
  size ($2b$) are set for both
  clouds according to their new properties. In particular,
    the grain charge is computed as in Paper I following
    \citet{2001ApJS..134..263W} for several dust sizes under conditions
    typical of a molecular cloud envelope, and an analytical
    expression is then obtained of the form:
    \begin{equation}
      Q \times 10^{4} [{\rm cm}^{1/2}~{\rm g}^{1/2}] =
      \frac{a}{m \times 10^{2} [{\rm g}] + b}
      \label{eq:dust_charges}
    \end{equation}
    where $a = -9.272 \times 10^{-3}$ $(-8.427 \times 10^{-4}$) and
    $b = 4.429\times 10^{-3}$ ($5.940 \times 10^{-7}$) for silicate
    (graphite) grains\footnote{This values have been obtained
      by fitting eq. \ref{eq:dust_charges} using the curve
      fitting tool \textit{cftool} in Matlab \citep{MATLAB:2020}.}.
    This analytic expression is used inside the code to
    automatically update the particle charge.
    The Coulomb rate is then
    computed from the new dust charge as in Paper I, and the
    cloud size depends on the updated cloud mass.

\end{itemize}

In Fig. \ref{fig:geometry} we present an schematic view
of both the coagulation and erosion processes for clarity.\par

According to this prescription, if we introduce several dust
populations of radii $a_{0} < a_{1} < ... < a_{n}$, particles
with $a_{d} > a_{n}$ may be formed, but those with
$a_{d} < a_{0}$ that should be created after an
erosion event will not be taken into account.
The implications of this
hypothesis on the physics are discussed later in Sec. \ref{sec:discussion}.

% ------------------------------------- %
% ------------------------------------- %
\section{Growth of a single-sized dust population}\label{sec:simu_turb}
% ------------------------------------- %
% ------------------------------------- %
In Paper I, we showed that filaments of charged dust are able
to form under conditions typical of a molecular cloud
envelope ($T = 6000$ K, $n_{\rm H} = 10$ cm$^{-3}$, ionization
fraction $\chi = 0.1$); these parameters are very similar
  to those for the warm neutral medium (WNM), so 
  this analysis is also valid for
  other diffuse regimes.
However, in Paper I dust particles were
not allowed to interact with each other and only suffered
from the drag of neutral gas and magnetic fields. \par

We have first tested the performance of the collision model
presented in the previous section using the results of
Paper I as input data. With that purpose, we have selected seven
regions where the dust-to-gas ratio is larger than the
mean expected value for the diffuse ISM and for consistency, we
keep the same names as in Paper I: H1, H2, H4, L1, L2, L4, and M1.
In regions H1-H4, the gas density value $\rho$ is greater than 1$\sigma$;
in regions L1-L4, it is lower than 1$\sigma$; and in region M1, the gas
density reaches intermediate values. In all cases, conspicuous filaments
of charged dust are formed aligned with the magnetic fields. We want to
note that the
computational domains for coagulation simulations are squared
domains that contain the original regions, so
their limits are slightly different as those in Paper I; the properties
of these regions are summarized in Table \ref{tab:regions}.\par

\begin{deluxetable*}{ccccccc}
\tablecaption{Properties of the selected regions\label{tab:regions}}
\tablewidth{0pt}
\tablehead{
\colhead{Region} & \colhead{Length} & \colhead{Resolution} & \colhead{$M_{\rm dust}$} & \colhead{$<\rho>$} & \colhead{$<B>$} & $t_{\rm lim}$ \\
\colhead{} & \colhead{pc} & \colhead{px}
& \colhead{$10^{29}$ g} & \colhead{$10^{-23}$ g cm$^{-3}$} & \colhead{$10^{-6}$ G} & \colhead{$10^{10}$ s} }
%\decimalcolnumbers
\startdata
H1 & 0.145 & 148 & 1.571 & 1.861 & 1.011 & 7.282 \\
H2 & 0.121 & 124 & 6.854 & 1.828 & 1.057 & 9.987  \\
H4 & 0.166 & 170 & 1.654 & 1.878 & 1.425 & 12.122  \\
L1 & 0.117 & 120 & 0.832 & 1.526 & 1.609 & 9.665  \\
L2 & 0.109 & 112 & 1.056 & 1.576 & 1.471 & 9.021  \\
L4 & 0.180 & 184 & 1.822 & 1.609 & 1.333 & 14.820  \\
M1 & 0.063 & 64 & 0.478 & 1.711 & 0.080 & 5.155  \\
\enddata
\tablecomments{Regions selected from the turbulence simulation
  (Paper I) to study the performance of the dust coagulation
  algorithm. In all cases, they are taken to be squared regions
  of a fixed length as given in the table, and with a uniform
resolution in x- and y- directions.}
\end{deluxetable*}

Given a region, we extract the gas properties and particle
positions, and set them inside a new domain with normalized limits
$[0, 1] \times [0, 1]$. The boundary conditions are chosen to be
flow out to ensure that the divergence-free constraint holds
$\nabla \cdot {\bf \vec{B}} = 0$. This implies that some mass of dust
will be inevitably lost, but we will take this fact into account
when discussing the performance of the algorithm.
The initial dust
population is homogeneous, with a common radius of $0.05~\mu$m, a charge
of $Z_{d} = -17$, and an internal solid density
$\rho_{d}^{\rm int} = 1$ g cm$^{-3}$. As explained in Sec.
\ref{sec:algorithm}, the computational test particles have to be
considered as the centroids of a squared cloud with linear size $2b$
that depends on the dust content. At the initial stage, all the clouds
have the same linear size, typically 1.4 times the pixel size; this
means that the clouds span a region of $2 \times 2$ pixels.
The system is then left to evolve
an arbitrary amount of time that differs from simulation to
simulation. The final time is subjectively set following the condition
that dust particles have moved enough to produce substantial coagulation,
but they have not been completely swept away from the domain. A typical
value is $t_{\rm lim} = 10^{11}$ s ($\sim 3 \times 10^{-3}$ Myr,
see Table \ref{tab:regions}) that, 
when compared with
the global simulation 
($t_{\rm lim}({\rm Paper~I}) = 1.37 \times 10^{13}$ s), is roughly a $7\%$ of
the total time. Hence, we expect that the phenomena observed in the
simulations here presented are representative of the evolution
of the global system (Paper I) at short time scales.
Besides, if we compare this time with the typical
  lifetime of a molecular cloud
  \citep[$\sim 10^{6}$ Myr,][]{2001ApJ...562..852H}
  it is evident that the timescales treated in this work
  are so small that we can consider that the final
  size distributions are representative for a molecular cloud at
  any stage of its evolution, provided that the ambient
  conditions (temperature, ionization fraction, magnetic field
  strength, and density)
are maintained.

% % % % % % % % % % % % % % % % % % % % % % % % % % 
\subsection{Evolution of the size distribution}\label{sec:turb_growth}
% % % % % % % % % % % % % % % % % % % % % % % % % % 
Starting from single-sized dust grains, and given the
fact that we are not allowing particles to decrease their
size, we do not expect to retrieve a realistic size distribution
of the type $dn/da \propto a^{q}$, $q=-3.5$
(\citealt{1977ApJ...217..425M}, hereafter MRN).
In fact, attempts to fit the dust size distribution to a
power law give very sharp values
for $q$ ($q \sim -4, -5$).\par

Globally, we find that a significant fraction
  of the dust population ($30 - 40 \%$) have a final
  radius greater than the initial value of $0.05~\mu$m, but
  only few of them (at most 2\%) acquires a size
  greater or equal than twice the initial value ($0.1~\mu$m). Besides,
  the growth also depends on the ambient conditions:
  particles that suffer from moderate growth (final sizes below $0.1~\mu$m) are found
  all over the domain, but the largest particles are only created
  in high dust density regions with weak magnetic fields; this
  is logical since we are working with charged dust particles
  with large charge-to-mass ratios, so they are effectively
  accelerated by the Lorentz force. At high density regions, however,
  the magnetic field strength decreases and the dust grains are
less accelerated. In every simulation we find one or two particles that are able to
acquire very large sizes (from $0.5$ to $1~\mu$m),
and they are initially located
in regions with a magnetic field strength lower than the mean.\par

\begin{figure*}
  \gridline{\fig{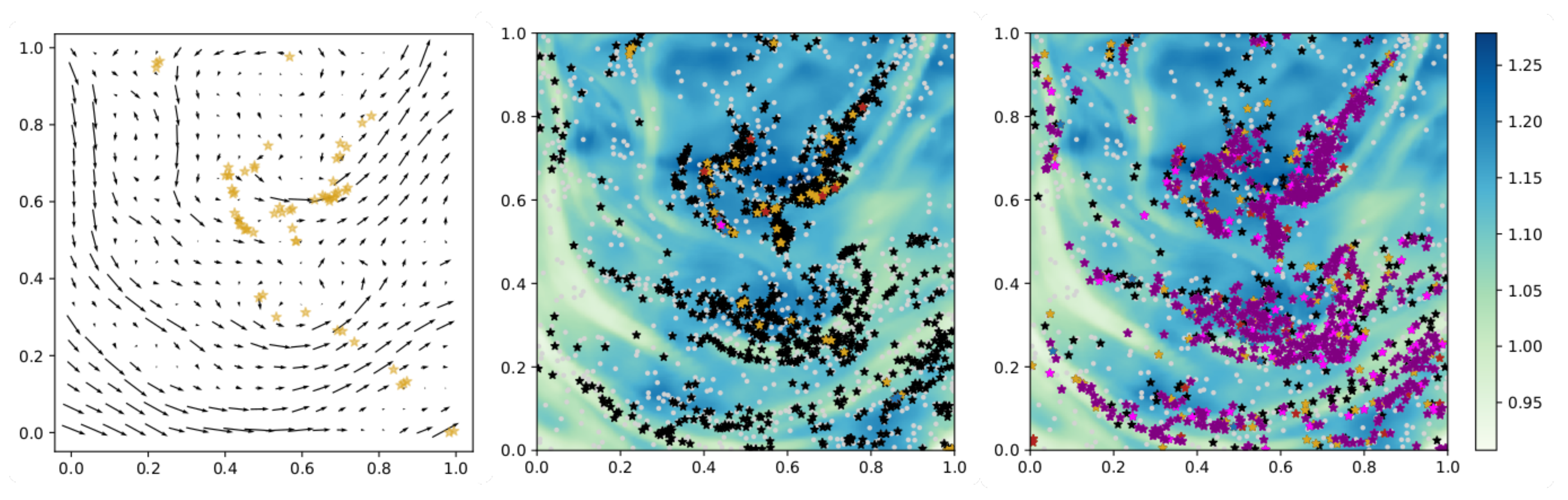}{\textwidth}{}}
  \caption{Initial distribution of test particles
    in the H1 region.
    \textit{Left}: particles that grow
    over $2a_{0} = 0.1~\mu$m (gold stars) plotted over
    the vectorial magnetic field. \textit{Middle}:
    initial distribution of test particles plotted over
    the gas density (in code units). Grey circles correspond to particles
    that do not experience growth, while stars
    correspond to those that increase their sizes. The
    color code of the stars correspond to the final
    size $a_{\rm fin}$: $a_{\rm fin} > a_{0}$ (black),
    $> 2a_{0}$ (gold), $> 4a_{0}$ (red), $> 6a_{0}$ (blue),
    $> 8a_{0}$ (magenta). \textit{Right}: same as the
    previous one, but the color code correspond to
    clouds of $a_{\rm fin} = 0.05~\mu$m
    that lose mass: $M_{\rm fin} < M_{0}$ (black),
    $< 10^{-1} M_{0}$ (gold), $< 10^{-2}M_{0}$ (red),
    $<10^{-3}M_{0}$ (blue), $< 10^{-4}M_{0}$ (magenta),
    and $< 10^{-5}M_{0}$ (purple).
    \label{fig:growth_singlesize}}
\end{figure*}

Finally, we have qualitatively explored the possible influence
of the morphology of the magnetic field in the observed dust growth.
With that purpose, we have plotted the initial positions of all
the particles that acquire a size greater than $0.1 ~\mu$m ($2a_{0}$)
over the vectorial magnetic field, and also over the
initial density field (see Fig. \ref{fig:growth_singlesize}). In general,
particle growth takes place inside the main
dust filaments, parallel to the magnetic field lines, but
the largest grains mentioned above form in high dust concentrations
located inside magnetic field loops. 

% % % % % % % % % % % % % % % % % % % % % % % % % % 
\subsection{Dust mass loss}\label{sec:turb_erosion}
% % % % % % % % % % % % % % % % % % % % % % % % % % 
According to our model, there are two possible outcomes for
a collision between two particles: one of them may grow, as we
have already seen, or if their relative velocity is larger
than the threshold, the less massive will suffer from erosion.
This erosion is one of the main mechanisms for dust mass
loss, and in practice we shall suppose that the mass lost after
an erosion is actually converted into smaller dust grains
that return to the gas phase and
are not followed by our algorithm.\par

First, we have analyzed how many clouds lose
dust mass but keep their original radius ($a_{d} = 0.05~\mu$m).
Almost half of the clouds are eroded and for approximately
50\% of them, the mass is reduced five orders of magnitude (see
Fig. \ref{fig:growth_singlesize} for an illustration).
This means that dust erosion is very likely to take
place in molecular cloud envelopes, preventing the dust grains
to acquire very large sizes, and increasing the abundance of
small dust particles that interact very effectively with
the ambient magnetic field.\par

We can further quantify the amount of dust that is lost in
the interactions, \textit{i.e.}, the dust mass that is
transferred to the smaller populations not tracked in our
simulation. Comparing the number of clouds at the final
stage of the simulation with the initial one, we can give a
lower limit for the dust mass that has gone out of the
domain, but that is not necessarily transferred
to the small-sized population. Since at $t = 0$ all the
clouds have the same mass, we conclude
that the amount of dust that has escaped the domain is
$\sim 5\%$ of the total initial dust mass. This means that
most of the mass, that is approximately a 40\% of the initial
one, is converted into small dust particles that
return to a gaseous state.

% ------------------------------------- %
% ------------------------------------- %
\section{Evolution of a realistic dust population}\label{sec:simu_real}
% ------------------------------------- %
% ------------------------------------- %
To complete the testing of the algorithm, we are
interested in seeing how a realistic dust size
distribution behaves. With that purpose, we have
generated some mock dust samples for the regions
studied in Sec. \ref{sec:simu_turb} and have
evolved them up to the same time limit. Our objective
is twofold: first, considering two independent dust populations
(one composed of silicates and another one of graphite
grains), study the variation in the power index
of the size distribution; and second, taking the
final dust size distributions, build the ultraviolet
extinction curve and quantify the differences that arise
due to dust coagulation and erosion events. In the
next section (Sec. \ref{sec:mock-sample}) we explain in
detail how we have generated the mock samples for these
tests, and the analysis of the size distribution is
performed in Sec. \ref{sec:mock-sizedistr}. Finally,
the study of the ultraviolet
extinction curve is presented in Sec. \ref{sec:mock-uvext}.

% % % % % % % % % % % % % % % % % % % % % % % % % % 
\subsection{Generation of a mock sample}\label{sec:mock-sample}
% % % % % % % % % % % % % % % % % % % % % % % % % % 
For a given region, we know the total amount of dust
as listed in Table \ref{tab:regions}. Then, we suppose that
the dust abundance is equally distributed between two populations,
one of silicate grains with solid density
$\rho_{\rm sil}^{\rm int} = 3.5$ g cm$^{-3}$ and another one
of graphite grains with $\rho_{\rm gra}^{\rm int} = 2.24$ g cm$^{-3}$.
Each population will be represented by ten dust families with sizes
ranging from $a_{min} = 50$~\AA~to $a_{max} = 0.25~\mu$m,
logarithmically spaced to have a better
sampling of the small end of the distribution:
5 nm, 7.7 nm, 11.9 nm, 18.4 nm,
28.4 nm, 43.9 nm, 67.9 nm, 104.8 nm,
161.9 nm,
250 nm. 
Since we are still working under conditions typical of a
molecular cloud envelope, we compute the grain charges and
Coulomb drag parameters for each dust
family and composition as in Paper I ($Z_{i}^{\rm sil}$, $Z_{i}^{\rm gra}$).
Individual
  simulations with identical initial conditions are carried out for
  the silicate and graphite populations for simplicity, because
  our algorithm do not allow interaction between
  particles from different populations.
\par

Independently on the dust abundance, for each region
and material we generate 1000
clouds, a quantity imposed by computational restrictions,
  and consider a homogeneous sampling of the populations, which
  implies that 100 clouds of each size family are followed. Then,
  these clouds are distributed across the domain
according to three criteria: particles with $a_{d} = 0.05~\mu$m
should follow the original dust distribution from Paper I, those
with $a_{d} = a_{min}$ will be placed according to the magnetic
field strength, and those with $a_{d} = a_{max}$ will be generated
according to the gas density; for $a_{min} < a_{d} < 0.05~\mu$m
and $0.05~\mu$m$< a_{d} < a_{max}$, we take a linear interpolation
between the boundary values. Each mock cloud is generated in the center of a computational cell, and its
velocity is also given as a function of the velocity field of the
original dust distribution, the Alfv\'en velocity at cell-centers, and
the gas velocity.
A detailed explanation of the methods
behind the random distribution of particles and the determination
of their velocities can be
found in Appendix \ref{appendix:mock}.\par

Finally, as we want this mock sample to be representative of
a realistic dust size distribution, we distribute the dust mass
among each population in order to satisfy the condition
$n_{i} = n_{min}(a_{i}/a_{min})^{-3.5}$.

% % % % % % % % % % % % % % % % % % % % % % % % % % 
\subsection{Variations in the size distribution}\label{sec:mock-sizedistr}
% % % % % % % % % % % % % % % % % % % % % % % % % % 
We have studied separately the behavior of
silicate and graphite grains, since differences are to be expected
due to the smaller value of $v_{crit}$ for graphites.\par

First, we analyzed the maximum achievable particle size.
In general, silicate particles acquire radii up to 
$\sim 0.34~\mu$m while graphite ones rarely grow beyond
the upper boundary of $a_{max} = 0.25~\mu$m; this is consistent
with the difference in the adopted values for $v_{crit}$. However,
there are some exceptions: in region L4, there is one silicate
grain that is able to grow up to $1.58~\mu$m starting from
a very small size ($a_{d} = 28.4$ nm). It is placed in a region
that coincides with one of the
filaments inside which dust grains are able to acquire
sizes greater than $0.2~\mu$m in the turbulent
simulation (Sec. \ref{sec:simu_turb}). For the graphite
grains, in region H2 there is one particle that grows
up to $a_{d} = 0.325~\mu$m, while in L1 another one reaches
$a_{d} = 0.356~\mu$m. In this case, however, both grains had an initial
radius of $a_{d} = 0.25~\mu$m, and therefore they are not
candidates to suffer erosion nor are accelerated by the magnetic fields
(they move with the gas). Although these variations in the size
distributions may seem relevant, we find very similar
values when fitting the populations to a MRN power law (see
Table \ref{tab:realistic_values}). In concordance with
the previous discussion, since silicate grains grow more
efficiently than graphite particles, the power index of the former population
is in general shallower than the latter one, but
they are always of the order of $q \sim -3.3$.\par

Then, we have again quantified the percentage of mass lost by
the system. We observe greater values than those of the
turbulent case (Sec. \ref{sec:simu_turb}) where the
mean was $\sim 35\%$ and approximately
half of that mass was reconverted into smaller
grains. This time, almost half of the
mass disappears (see Table \ref{tab:realistic_values}), but
it is not distributed among the smaller population still
present in the domain: practically 80\% of the mass lost
in this simulations is carried by very small particles
of sizes 5 nm and 7 nm that rapidly drifts away from the domain.
Since we are assuming a MRN-like initial size distribution, and
we start with 100 particles for each radius, the smaller ones
represent more mass than the larger ones. Besides, due to their
very small sizes, their dynamics is governed by the charges, so
they are effectively accelerated by the magnetic field, barely
feel the gas, and move away. The effect is more pronounced for
the graphite population because their charge parameters are
slightly larger (see e.g. \citealt{2004tcu..conf..213D}).

\begin{deluxetable*}{cccccccc}
\tablecaption{Size distribution properties for the realistic simulations\label{tab:realistic_values}}
\tablewidth{0pt}
\tablehead{
  \colhead{Region} & \colhead{$a_{max}^{\rm sil}$} & \colhead{$a_{max}^{\rm gra}$} & \colhead{$q^{\rm sil}$} & \colhead{$q^{\rm gra}$} & \colhead{$M_{\rm lost}^{\rm sil}$} & \colhead{$M_{\rm lost}^{\rm gra}$} & \colhead{$b_{\rm FUV}/b_{\rm FUV}^{\rm MRN}$}\\
\colhead{} & \colhead{$\mu$m} &  \colhead{$\mu$m} & \colhead{} & \colhead{} & \colhead{\%} & \colhead{\%}&  \colhead{}}
%\decimalcolnumbers
\startdata
H1 & 0.338 & 0.250 & -3.31 & -3.33 & 45.56 & 50.28 & 0.76\\
H2 & 0.338 & 0.325 & -3.33 & -3.39 & 48.83 & 50.03 & 0.79\\
H4 & 0.332 & 0.250 & -3.28 & -3.35 & 45.51 & 54.30 & 0.75\\
L1 & 0.358 & 0.356 & -3.28 & -3.40 & 47.69 & 52.57 & 0.78\\
L2 & 0.338 & 0.250 & -3.26 & -3.36 & 44.97 & 49.13 & 0.75\\
L4 & 1.580 & 0.250 & -3.39 & -3.33 & 48.89 & 54.33 & 0.86\\
M1 & 0.352 & 0.250 & -3.07 & -3.11 & 61.33 & 65.79 & 0.55\\
\enddata
\tablecomments{Maximum grain sizes, power index of the size distribution, and percentage of
  mass lost for silicate and graphite particles in the realistic simulations.
  The last column also shows the ratio between the FUV slope of the given
curves and the initial MRN distribution.}
\end{deluxetable*}

% % % % % % % % % % % % % % % % % % % % % % % % % % 
\subsection{Variations in the UV extinction curve}\label{sec:mock-uvext}
% % % % % % % % % % % % % % % % % % % % % % % % % % 
Motivated by the deviations of the size distributions from
the initial MRN shape, we have investigated the variations
in the UV extinction curve that may arise due to
grain growth in molecular cloud envelopes.\par

We follow the approach by \citet{2013ApJ...770...27N} and
compute the extinction curve as:

\begin{equation}
  A_{\lambda} = 1.086\sum_{j}\int_{a_{min}}^{a_{max}} \pi a^{2}Q_{\lambda,j}^{\rm ext}(a)n_{j}(a) da
  \label{eq:ext_curve}
\end{equation}

We take $a_{min} = 50$\AA, $a_{max} = 0.25~\mu$m, and
$n_{j}(a) = K_{j}n_{H}a^{q}$ \citep{2001ApJ...548..296W},
where the index $j$ denotes the grain species (graphite
or silicate) and $K_{\rm sil} = 10^{-25.11}$ cm$^{2.5}$,
$K_{\rm gra} = 10^{-25.13}$ cm$^{2.5}$. The factor
$Q_{\lambda, j}$ in equation \ref{eq:ext_curve} is the
extinction efficiency factor for species $j$, and is
computed from the Mie theory using the
optical constants\footnote{The optical constants
are available at Draine's web page: \url{https://www.astro.princeton.edu/~draine/dust/dust.diel.html}.} by \citet{2003ApJ...598.1026D}. For
graphite grains, we follow the standard 1/3 - 2/3 approach
($Q_{\lambda,{\rm gra}} = (Q^{\parallel}_{\lambda,{\rm gra}} + 2Q^{\perp}_{\lambda,{\rm gra}})/3$, 
\citealt{2001ApJ...548..296W,2013ApJ...770...27N})
and take the dielectric constants for particles
of radius $0.01~\mu$m.\par

\begin{figure}[h!]
  \plotone{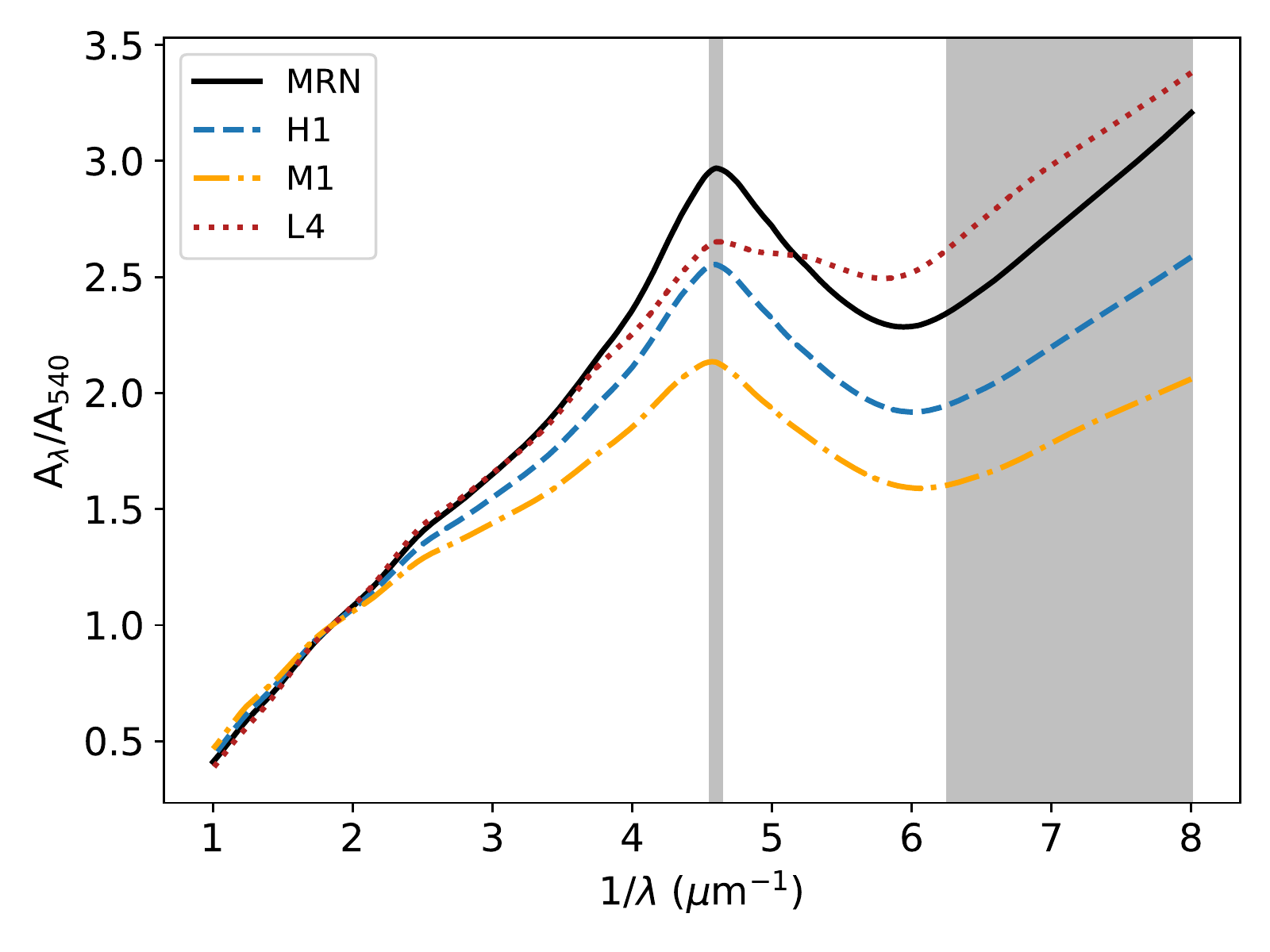}
  \caption{Normalized synthetic extinction curve for regions H1
    (blue dashed line), M1 (orange dashed-dotted line), L4
    (red dotted line),
    and for the initial MRN size distribution
    (black solid line). The shadowed region corresponds to
    the FUV part of the curve, where we perform a linear fitting
    of the slope. The vertical gray line marks the position
  of the 2175 \AA~bump.}
  \label{fig:sample_extcurve}
\end{figure}

For each region, we have built the extinction
curve from equation \ref{eq:ext_curve}
taking the values of $q$ listed in Table \ref{tab:realistic_values}.
We have further normalized the curve dividing by
the extinction at 540 nm, which is the effective wavelength
of the V band; an example is shown
in Fig. \ref{fig:sample_extcurve}. Using these normalized curves, we can now
proceed to study the variations that arise at ultraviolet wavelengths
due to grain growth. \par

As we mentioned in the introduction, there are two main features
of the ultraviolet extinction curve that may vary due
to grain growth. The most prominent one is the bump at 2175 \AA~(see
Fig. \ref{fig:sample_extcurve}), but other species
apart from
graphite and silicate grains should be considered for a realistic
treatment, especially PAHs. However, we want to note that
the
very small graphite grains here considered have typical sizes
similar to those of PAHs (of the order of a few nm).
 The main difference between graphite
  and PAHs is the nature of the chemical bonds (C-H in PAHs that
  are not present in graphite grains) that affect the dielectric
  constants and, in consequence, the shape of the UV bump.
We will consider to include
PAHs for future applications, but in this work we only aim to
demonstrate the feasibility of our method, so we will
restrict to the study of the FUV slope.\par

Comparing the values in Table \ref{tab:realistic_values}
and the previous discussions on grain growth, we see
that even a modest growth is translated into
variations in the FUV slope of the order of $\sim 20\%$.
A depletion of small particles, which is translated into
a shallower value of the power index of the size
distribution $(|q| < 3.5)$, is always accompanied by a decrease
of the FUV slope. Two regions stand out from the rest:
L4 shows less conspicuous variations in the FUV slope, reaching a
ratio $b_{\rm FUV}/b_{\rm FUV}^{\rm MRN} = 0.86$; actually, it is
straightforward from values in Table \ref{tab:realistic_values} that
this is the only region where the small graphite population
has been less depleted than the silicate one.  For M1 the
effects are more acute: the value of the slope is reduced
to a half of the MRN initial distribution due to grain escape in
the simulations, but also due to the destruction of small particles
during erosion events: only a 26\% (silicate) and a 30.5\% (graphite) of the
lost mass is carried out by clouds that escape from the domain.\par

We have finally assessed which of the two populations
(silicate or graphite) exerts more influence in the
FUV extinction curve. With that purpose, we have used the values
above presented together with those from mock populations
with $q$ ranging from -3.0 to -3.5. As a general rule, for a
given $q^{\rm gra}$, there are not significant variations for
the $b_{\rm FUV}/b_{\rm FUV}^{\rm MRN}$ ratio for $q^{\rm sil} \geq q^{\rm gra}$.
However, we do observe significant variations when we fix
the power index of the silicate size distribution and take
$q^{\rm gra} \geq q^{\rm sil}$. From this, we conclude that
variations in the FUV slope of the UV extinction curve
mainly arise from depletion of small graphite grains
that return to the gas phase.

% ------------------------------------- %
% ------------------------------------- %
\section{Discussion and conclusions}\label{sec:discussion}
% ------------------------------------- %
% ------------------------------------- %
In this work, we have studied the growth and erosion
of interstellar dust grains under conditions
typical of a molecular cloud envelope for different
populations based on their evolution
in 2D MHD simulations. Although our methodology is
different from that in the common literature, we
are going to compare
our results with those by
\citet{2004ApJ...616..895Y} (hereafter YLD04)
and by HY09
to determine 
the advantages of a direct dust modeling
over a more general statistical (that may be regarded
as 3D)
approach.
Both works consider charged dust grains
accelerated by MHD turbulence in several
phases of the ISM, and only take
into account coagulation and shattering
effects. We will discuss only the
results for the WNM, since its properties
are essentially the same as those adopted for
our molecular cloud envelope model (see
Sec. \ref{sec:simu_turb}).\par

YLD04 adopted a statistical description of MHD
turbulence and damping processes to derive
the grain velocities in the WNM and to estimate the
critical size for silicate and carbonaceous grains
in order to experience shattering and coagulation.
According to their study, only grains larger than
0.2 $\mu$m are able to suffer from shattering, while
coagulation dominates for sizes lower than 0.02 $\mu$m for
silicates, and 0.04 $\mu$m for carbonaceous grains. Similar
results were reported by HY09, since they adopted
YLD04's velocities in order to evolve the grain size distribution.
According to the latter work, the maximum grain size $a_{max} = 0.25~\mu$m 
cannot be superseded because either coagulation is not
efficient enough or because they would be destroyed by shattering
effects. However, the results presented in Sec. \ref{sec:simu_turb}
and in Table \ref{tab:realistic_values}
are not in agreement with that statement.
In Table \ref{tab:turb_eros_stats}, we show the number of coagulation
and erosion events for each simulation, as well as their mean
duration time. 
Focusing on the turbulent simulation
(Sec. \ref{sec:simu_turb}), it is straightforward
that even a single-sized dust distribution
may experience several coagulation and shattering events, but
that their relative frequency differs. For regions H1, H2, and L2
coagulation and shattering are essentially balanced, while for
H4, L1, L4, and M1 shattering dominates. However, since
the mean and median interaction times are, in general, greater
for shattering, the growing mechanisms for small dust particles
are countered. These conclusions are further sustained by
the results for the realistic size distributions considered
in Sec. \ref{sec:simu_real}, since for them the shattering
events are approximately five times more common than coagulation ones
for both silicate and graphite grains. Nevertheless,
we do observe growth for the silicate population due to the shorter
timescales for erosion, while graphite grains barely supersede the
maximum initial size due to the low potential barrier
for erosion (lower value of $v_{crit}$).
We want to emphasize that we do not find any common trends between
  regions with the same gas density properties, \textit{i.e.}, coagulation
  and shattering are not governed by the gas but rather by the local
  morphology of the magnetic field. 
\par

\begin{deluxetable*}{cccccccc}
\tablecaption{Coagulation and shattering parameters\label{tab:turb_eros_stats}}
\tablewidth{0pt}
\tablehead{
  \colhead{} & \colhead{H1} & \colhead{H2} & \colhead{H4} & \colhead{L1} & \colhead{L2} & \colhead{L4} & \colhead{M1}
}
%\decimalcolnumbers
\startdata
% Turbulent simulation
& & Turbulent & simulation \\
N$_{\rm coag}$ & 2664 & 895 & 1725 & 959 & 1635 & 1841 & 577 \\
$<t_{\rm coag}> $ ($10^{9}$ s) & 0.71 & 1.45 & 1.42 & 1.39 & 1.70 & 2.24& 0.23\\
median $t_{\rm coag}$ ($10^{7}$ s) & 11.3 & 29.9 & 2.05 & 6.44 & 4.51 &
3.21 & 1.17 \\
N$_{\rm shat}$ & 2680 & 803 & 2682 & 1341 & 1592 & 2695 & 1190 \\
$<t_{\rm shat}>$ ($10^{9}$ s) &  1.11 & 1.78 & 1.25 & 1.18 & 1.26 & 1.70 & 0.82\\
median $t_{\rm shat}$ ($10^{8}$ s) & 1.21 & 3.53 & 1.16 & 1.40 & 1.70 &
1.16 & 0.59\\
& & Realistic: & silicate \\
N$_{\rm coag}$ & 174 & 144 & 149 & 160 & 179 & 154 & 221\\
$<t_{\rm coag}> $ ($10^{9}$ s) & 0.54 & 0.65 & 1.00 & 1.20 & 1.75 &
1.43 & 0.13 \\
median $t_{\rm coag}$ ($10^{7}$ s) & 2.59 & 2.10 & 3.83 & 1.19 & 5.11 &
1.38 & 1.15\\
N$_{\rm shat}$ & 772 & 878 & 805 & 883 & 979 & 873 & 1113\\
$<t_{\rm shat}>$ ($10^{8}$ s)  & 7.57 & 4.09 & 5.07 & 4.74 & 4.03 &
6.22 & 6.00\\
median $t_{\rm shat}$ ($10^{7}$ s) & 5.24 & 1.60 & 1.76 & 1.42 & 1.19 &
2.31 & 9.71\\
& & Realistic: & graphite \\
N$_{\rm coag}$ & 165 & 144 & 136 & 155 & 189 & 150 & 167\\
$<t_{\rm coag}> $ ($10^{8}$ s) & 4.4 & 5.36 & 5.8 & 4.06 & 14.30 & 8.94 & 0.43\\
median $t_{\rm coag}$ ($10^{7}$ s) & 1.13 & 0.81 & 2.16 & 0.33 & 1.76 &
0.64 & 0.38\\
N$_{\rm shat}$ & 908 & 943 & 1068 & 1039 & 1041 & 1024 & 1259\\
$<t_{\rm shat}>$ ($10^{8}$ s)  & 8.23 & 5.13 & 6.94 & 5.99 & 6.56 &
9.48 & 7.25\\
median $t_{\rm shat}$ ($10^{7}$ s) & 3.42 & 1.91 & 2.07 & 1.28 & 1.32 &
1.80 & 7.44\\
\enddata
\tablecomments{Coagulation and shattering parameters for the
  different simulations: turbulent (Sec. \ref{sec:simu_turb}),
  and realistic populations of silicate and graphite grains
(Sec. \ref{sec:simu_real}).}
\end{deluxetable*}

The origin of the discrepancies between the work here presented
and the ones by YLD04 and HY09 is very likely the fundamentally
different followed approaches. While YLD04 and HY09 adopt a
statistical 3D treatment for the grain's velocities, we have studied
dust interactions inside dense filaments formed in 2D
MHD simulations, so
the collision frequencies in our work are consequently greater
because dust particles in our simulations are
  restricted to move in a plane; introducing one additional
  degree of freedom will likely reduce the collision rates, although
the filamentary distribution will always favor grain-grain collisions.
Then, when
studying the general behavior of dust grains in a turbulent, generic
ISM medium, YLD04 and HY09's approaches may be considered a fair
approximation, while for more specific studies, such as the one
here presented, it may be useful to evolve a full realistic dust
population submitted to the resolved forces of the gas and magnetic field.\par

Finally, we want to discuss the physical consequences
  of the hypothesis that the mass of dust lost in erosion
  events returns to the gas phase. When dealing with
  grain shattering, it is customary to assume that the dust mass
  is redistributed among the lower-sized population (see e.g. HY09).
  Although this approach is reasonable when evolving a dust
  size distribution, its implementation inside a framework
  where MHD codes are involved is not trivial. Trying to increase the small
  sized population of neighboring particles would inevitably
  introduce more restrictions to the algorithm difficult to
  justify and that would obscure the interpretation of the
  results. On the other hand, the assumption that the
  residual dust mass lost after a collision is converted into
  nanometer-sized particles that may return to the gas phase
  is more natural and easier to interpret. \par

To sum up, in this paper we have presented a new formulation
to study the variations in the grain size distribution using
a particle-in-cell code coupled with a MHD code. We have implemented
this formulation in the MHD code Athena, and have used it
to study the variations in the dust size distribution in a molecular
cloud envelope and the expected variations in the FUV extinction
curve. This work is very related to other recent attempts to include
the evolution of the dust size distribution in MHD codes
\citep{2018MNRAS.478.2851M,2018ApJ...863...97T,2020ApJ...903..148L}
but it is the only one that considers charged dust particles and
that is designed for studies of the diffuse ISM. Currently, and up
  to the knowledge of the authors, the other numerical code that is
  considering the evolution of charged dust grains in the
  ISM is GIZMO \citep{2015MNRAS.450...53H,2017MNRAS.469.3532L}, but
  the public version only takes into account the interaction of dust
  particles with the gas and the magnetic fields.

% ------------------------------------- %
% ------------------------------------- %
% ACKNOWLEDGEMENTS
\acknowledgments

We want to thank an anonymous referee for their useful suggestions
that have helped to improve the clarity of this manuscript.
We want to thank Juan Carlos Vallejo for fruitful discussion
about the coagulation algorithm.
L. B.-A. acknowledges Universidad Complutense de Madrid and Banco
Santander for the grant ``Personal Investigador en
Formaci\'on CT17/17-CT18/17''. This work has been partially funded by the
Ministry of Economy and Competitiveness of Spain
through grants MINECO-ESP2015-68908-R and MINECO-ESP2017-87813-R.
This
research
has made use of NASA’s Astrophysics Data System.

\appendix

\section{Generation of a mock sample} \label{appendix:mock}
In order to generate the position and velocity
of a mock cloud of particles with radius $a_{i}$,
$a_{min} \leq a_{i} \leq a_{max}$, we make use
of three probability
matrices. We take as input data the matrix of magnetic
field modulus in the region of interest $M_{\rm B}$, the
matrix of the gas velocity modulus $M_{\rm gas}$, and
the matrix of dust mass $M_{\rm d}$ from Paper I; we normalize
these matrices so that the maximum value is 1. Then, for
every cloud of particles with radius $a_{i}$, the associated
probability matrix $M_{\rm prob}$ is computed as follows:

\begin{eqnarray}
  M_{\rm prob} =  |A| M_{\rm B} + (1 - |A|)M_{d}, ~~~ A = \frac{a_{0} - a_{i}}{a_{0} - a_{min}}, ~~~~~ \text{if} ~a_{d} \leq a_{0} \label{eq:mprob_1}\\
  M_{\rm prob} = |C| M_{\rm gas} + (1 - |C|)M_{d}, ~~ C = \frac{a_{i} - a_{0}}{a_{max} - a_{0}}, ~~~~~~\text{if} ~a_{d} > a_{0} \label{eq:mprob_2}
\end{eqnarray}

where $a_{0} = 0.05~\mu$m is the adopted radius for the simulations in Paper I.\par

The probability matrix $M_{\rm prob}$ gives the 2D probability distribution
of the clouds with radius $a_{i}$.
To assign a random position following this
distribution, we marginalize $M_{\rm prob}$ over $y$ and retrieve the cumulative
distribution function for the $x$-position. Then, we take a random
number from a uniform distribution and retrieve the associated $x$-position
from the cumulative distribution. Finally, the $y$-position is generated
in a similar manner, but this time the marginalized distribution is
imposed by the $x$-coordinate (it is a conditioned probability).\par

The velocity of the cloud is set based on its position in the grid.
Since we are placing the clouds at the center of the
computational cells, we can apply a linear relationship analogous to
\ref{eq:mprob_1} and \ref{eq:mprob_2} but substituting the probability
matrices $M_{\rm B}$, $M_{\rm gas}$, and $M_{d}$ by the corresponding
values of the velocities at that cell, ${\bf v}_{A}$, ${\bf v}_{\rm gas}$,
and ${\bf v}_{d}$;
this is done on a component-by-component basis.

% BIBLIOGRAPHY - At the very end

\bibliography{references}{}
\bibliographystyle{aasjournal}



\end{document}